\documentclass[aps,showpacs,manuscript,12pt]{revtex4}
\usepackage{amssymb}
\usepackage{amsmath}
\usepackage{graphicx}

\begin{document}
\title{\textbf{Generalized Grad-Shafranov equation for gravitational Hall-MHD
equilibria$^{\S }$}}
\author{C. Cremaschini$^{a}$, A. Beklemishev$^{b}$, J. Miller$^{c,d}$ and M.
 Tessarotto$^{e,f}$}
\affiliation{\ $^{a}$Department of Astronomy, University of
Trieste, Italy, $^{b}$ Budker Institute of Nuclear Physics,
Novosibirsk, Russia, $^{c}$International School for Advanced
Studies, SISSA and INFN, Trieste,  Italy,  $^{d}$Department of
Physics (Astrophysics), University of Oxford, Oxford, U.K.,
$^{e}$Department of Mathematics and Informatics, University of
Trieste, Italy, $^{f}$Consortium of Magneto-fluid-dynamics,
University of Trieste, Italy}
\begin{abstract} The consistent theoretical description of
gravitational Hall-MHD (G-Hall-MHD) equilibria is of fundamental
importance for understanding the phenomenology of accretion disks
(AD) around compact objects (black holes, neutron stars, etc.).
The very existence of these equilibria is actually suggested by
observations, which show evidence of quiescent, and essentially
non-relativistic, AD plasmas close to compact stars, thus
indicating that accretion disks may be characterized by slowly
varying EM and fluid fields. These (EM) fields, in particular the
electric field, may locally be extremely intense, so that AD
plasmas are likely to be locally non-neutral and therefore
characterized by the presence of Hall currents. This suggests
therefore that such equilibria should be described in the
framework of the Hall-MHD theory. Extending previous approaches,
holding for non-rotating plasmas or based on specialized
single-species model equilibria which ignore the effect of
space-time curvature, the purpose of this work is the formulation
of a generalized Grad-Shafranov (GGS) equation suitable for the
investigation of G-Hall-MHD equilibria in AD's where
non-relativistic plasmas are present. For this purpose the
equilibria are assumed to be generated by a strong axisymmetric
stellar magnetic field and by the gravitating plasma
characterizing the AD.
\end{abstract}

\pacs{95.30.Qd,97.10.Gz}
\date{\today }
\maketitle



\section{Introduction and motivations}

This work is part of a research investigation of accretion disks
arising around compact objects (black holes, neutron stars, etc.
). In this first paper we formulate a generalized Grad-Shafranov
equation which determines the self-generated equilibrium magnetic
field, via a quiescent dynamo effect.

The occurrence of dynamo effects of this type is essentially
ubiquitous in character both in laboratory and space plasmas. In
particular, it is well known that the phenomenon may exist even in
plasmas which are at fluid (or kinetic) equilibrium, or
sufficiently close to it, in the sense that the fluid fields (and,
respectively, the kinetic distribution function) characterizing
the plasma are stationary (or quasi-stationary). This corresponds
to the appearance of a self-generated magnetic field in quiescent
plasmas, i.e., in the absence of significant turbulence phenomena.
The same may happen also in accretion disk (AD) plasmas. In
particular, the self-generated magnetic field may become
comparable in magnitude, or even larger, than that may be produced
by the central object. This circumstance, it should be stressed,
might occur even in the case of low density AD plasmas. In this
paper we intend to pose the problem of the theoretical description
of self-consistent gravitational plasma equilibria appropriate for
modelling AD plasmas (in this context, what is meant by an
"equilibrium" is stationary-flow solution). Their investigation is
of fundamental importance for understanding the phenomenology of
accretion disks. The very existence of these equilibria is
actually suggested by observations, which show evidence of
quiescent and essentially non-relativistic, AD plasmas close to
compact stars, thus indicating that accretion disks may be
characterized by slowly varying EM and fluid fields. Gravitational
Hall-MHD equilibria (G-Hall-MHD) represent an important aspect of
the phenomenology accretion disk (AD) plasmas. This is true
especially for those AD which occur near compact objects (black
holes, neutron stars, etc. ) where the presence of strong EM
fields is expected. These fields, and particularly the electric
field, may locally be extremely intense, so that AD plasmas are
likely to be locally non-neutral and therefore characterized by
the presence of Hall currents. In this paper we analyze the
combined role of the diamagnetic effects arising in a rotating
plasma, Hall currents produced by local deviations from charge
neutrality, and the influence of space-time curvature produced by
the central object. It is important to stress, however, that two
classes of equilibria can in principle be distinguished, which are
respectively \textit{fluid} and also \textit{kinetic}. In the
first case only the fluid fields are required to be stationary,
while in the second also the kinetic distribution function must be
stationary. In addition, for kinetic equilibria, additional
prescriptions can in principle be added. Therefore, the two
problems, although evidently related, should be treated
separately. The purpose of this work is the investigation of the
first type of equilibrium. In particular, here we intend to
address the problem of the formulation of a generalized
Grad-Shafranov (GGS) equation suitable for the investigation of
G-Hall-MHD equilibria in AD's. The investigation of kinetic
equilibria is instead left to the accompanying paper
\cite{Cremaschini2008b}. Based on a two-fluid description of the
plasma, in which relativistic corrections due to space-time
curvature are taken into account only by introducing a suitable
Pseudo-Newtonian potential, the theory is applicable to the
investigation of axisymmetric equilibria which occur in the
central regions of the AD where the plasma particles are taken to
still be non relativistic in the sense that their rotational
velocities are much smaller that $c$. In the same region the
plasma flow velocity and current density of each plasma species
are determined self-consistently by taking into account the
relevant diamagnetic currents, including the Hall current and the
gravitational drift produced by space-time curvature. The theory
permits the explicit numerical determination of the equilibrium
magnetic field. The approach appears to be relevant for the
investigation of a variety of possible plasma equilibria in AD's.
Despite earlier investigations, carried out by several authors,
including for example Chandrasekhar (1956, \cite{Chandrasekhar})
and Morozov and Solov'ev (1963, \cite{Morozov}) and more recently
by Krasheninnikov and Catto \cite{Catto1999}, Throumoulopoulos and
Tasso (2001) \cite{Tasso}, McClements and A. Thyagaraya (2001,
\cite{McClements}) and by Ghanbari and Abbassi (2004,
\cite{Ghanbari}), a consistent theoretical description of these
equilibria - a challange for astrophysicists and mathematical
physicists alike - has not yet been obtained. We refer in
particular both to the approximate solutions and to the nature of
the physical models adopted so far. For example, in some of
previous works, factorized
approximate solutions for the dipolar magnetic field were used \cite%
{Catto1999,Tasso}, while in many cases the AD model is based on a
single-fluid ideal-MHD description, assuming quasi-neutral
rotating or non-rotating plasmas (see for example
\cite{Catto1999}) which are assumed to be subject to the action of
a purely classical Newtonian potential for the central object.
Other treatments \cite{Ghanbari}, which investigate the problem of
self-gravitating accretion disks, adopt a self-similar solution
method while ignoring completely relativistic effects and the
possible presence of self-generated EM fields. In a subsequent
development McClements and Thyagaraja \cite{McClements} tackled
the problem of constructing the relevant set of fluid equations,
in particular a generalized Grad-Shafranov equation for the
magnetic field, to be applied to numerical simulations of
gravitational AD plasmas. \ Their approach, while still
considering only a purely non-relativistic quasi-neutral plasma,
represents an interesting development. In fact, adopting a more
realistic two-fluid model based on ideal-MHD equations, it allows
- in principle - the numerical investigation of gravitational
plasma equilibria with arbitrary flows and subject to the action
of an arbitrary self-consistent (axisymmetric) magnetic field.
Nevertheless several aspects of the theory deserve further
investigation. \ A first basic issue is related to the very
formulation of the theoretical model appropriate for AD plasmas.
In fact, AD's close to compact objects can be characterized by
extremely intense EM (both magnetic and electric) fields as well
by gravitational fields. It follows that AD plasmas are likely to
be locally non-neutral and therefore Hall currents may be present.
This suggests therefore that such equilibria should be described
in the framework of the Hall-MHD theory. These equilibria are here
denoted as gravitational Hall-MHD (G-Hall-MHD) equilibria. \ In
addition, for the description of equilibria occurring close to
compact stars, the effect of space-time curvature is expected to
become significant. Another problem to be further clarified,
however, is related to the determination of the self-generated
toroidal magnetic field arising in AD plasmas. In fact at
equilibrium, in the absence of radial flows and of an
externally-generated toroidal magnetic field, the self-consistent
magnetic field is expected to be purely poloidal. Its precise
relationship with the poloidal magnetic field in the plasma as
well with the relevant diamagnetic currents driven by plasma
inhomogeneities and flows has yet to be determined. Finally, an
interesting theoretical issue is related to the possible existence
of bifurcated plasma equilibria. These might correspond to AD's
characterized - in some suitable sense - by the presence of "high"
or "low" magnetic fields respectively. Besides the existence of
the bifurcation effect, an interesting closely related question is
obviously its possible role in the determination of the observed
phenomenology of AD's occurring close to compact objects. In this
paper, the focus of the investigation is on the combined role of
the diamagnetic effects produced by the rotating plasma, and in
particular the influence of Hall current and of space-time
curvature, for the possible generation of \textit{quiescent dynamo
effects} in these equilibria. This permits a detailed analysis of
the formation and the structure of these equilibria occurring in
the central regions of the AD where non-relativistic plasmas are
actually expected. Extending previous approaches, (in particular
Krashennikov and Catto, 1999 \cite{Catto1999} and Throumoulopoulos
and
Tasso, 2001 \cite{Tasso} and McClements and Thyagaraja, 2001 \cite%
{McClements}), the main goal of this work is the formulation of a
generalized Grad-Shafranov (GGS) equation suitable for the
investigation of G-Hall-MHD equilibria in AD's where
non-relativistic plasmas are present (Cremaschini et al., 2008
\cite{Cremaschini2008}). For this purpose the equilibria are
assumed to be generated by a strong axisymmetric stellar magnetic
field and by the gravitating plasma characterizing the AD. Basic
features of the theoretical model adopted include: the assumptions
of finite plasma rotation, two-species fluid fields,
divergence-free electric current density and (primarily) toroidal
plasma current. As a consequence, an equilibrium equation is
obtained from Ampere's law for the poloidal magnetic flux
function. An approximate solution method for the GGS equation is
presented which allows one to construct systematically approximate
analytical solutions near the equatorial plane of the AD. The
approach may be relevant for the investigation of a variety of
possible G-Hall-MHD equilibria in AD's.

\section{Construction of gravitational-Hall-MHD equilibria}

Let us now assume that the AD plasma is formed by two species of
charged particles: one species of ions and one of electrons. The
plasma is considered stationary, "collisionless" (or, in a proper
sense, weakly collisional) and axisymmetric, i.e., independent of
the azimuth $\varphi $ when referred to a set of cylindrical
coordinates $(R,\varphi ,z)$ for which the axis $R=0$ is a
principal axis. In such a case a complete description of
the plasma can be obtained in terms of the species-dependent number density (%
$n_{s}$), flow velocity ($\mathbf{V}_{s}$) and tensor-pressure ($\underline{%
\underline{\Pi }}_{s}$), together with the EM fields $\left\{ \mathbf{E}%
=-\nabla \phi \mathbf{,B}\right\} ,$ both externally produced
and/or self-generated, and the effective gravitational potential
produced by the central object and by the plasma itself, in which
the plasma is immersed. In particular, denoting by $\mathbf{b}$
the local unit vector of the magnetic field $\mathbf{B}$, ignoring
electron inertia contributions and the effect of binary
collisions, the species momentum balance equations read
respectively for ions and electrons%
\begin{equation}
m_{i}n_{i}\mathbf{V}_{i}\cdot \nabla \mathbf{V}_{i}+\nabla \cdot \underline{%
\underline{\Pi }}_{i}+m_{i}n_{i}\nabla _{R}U^{eff}-q_{s}n_{i}\mathbf{E}-%
\frac{q_{i}}{c}n_{i}\mathbf{V}_{i}\times \mathbf{B=0,}
\label{momentum-i}
\end{equation}%
\begin{equation}
\nabla \cdot \underline{\underline{\Pi }}_{e}-q_{e}n_{e}\mathbf{E}-\frac{%
q_{e}}{c}n_{e}\mathbf{V}_{e}\times \mathbf{B=0.}
\label{momentum-e}
\end{equation}%
It follows that the plasma current density takes the form $\mathbf{J=J}%
_{\parallel }\mathbf{b+J}_{\perp H}+\mathbf{J}_{\perp D},$ where $\mathbf{J}%
_{\parallel }$ is the parallel current density to be defined so as
to fulfill identically the isochoricity condition $\nabla \cdot
\mathbf{J=0.}$
Here $\mathbf{J}_{\perp H}=\frac{c\rho \mathbf{E\times b}}{B}$ and $\mathbf{J%
}_{\perp D}=\frac{c\rho \mathbf{E}^{(d)}\mathbf{\times b}}{B}$
denote
respectively the so-called Hall and diamagnetic current densities, while $%
\rho $ and $\mathbf{E}^{(d)}$ are the local charge density of the
plasma and the diamagnetic electric field $\mathbf{E}^{(d)}$ at
equilibrium, which is defined so that
\begin{equation}
\rho \mathbf{E}^{(d)}\equiv -m_{i}n_{i}\mathbf{V}_{i}\cdot \nabla \mathbf{V}%
_{i}-\nabla \cdot \underline{\underline{\Pi }}-m_{i}n_{i}\nabla
_{R}U^{eff}. \label{Diamagnetic E_field}
\end{equation}%
Here $U^{eff}$ denotes a suitable effective (pseudo-Newtonian)
gravitational potential taking into account relativistic
corrections of the Newtonian potential \cite{Cremaschini2008}.
Moreover, we assume that for a strongly
magnetized plasma, the tensor pressure $\underline{\underline{\Pi }}%
=\sum\limits_{s=e,i}$ $\underline{\underline{\Pi }}_{s}$ is
diagonal,
i.e., $\underline{\underline{\Pi }}_{s}$ is of the form $\underline{%
\underline{\Pi }}_{s}=P_{s\parallel }\mathbf{bb}+P_{s}\left( \underline{%
\underline{\mathbf{1}}}-\mathbf{bb}\right) $, where $P_{s\parallel }$ and $%
P_{s}$\ denote respectively the scalar parallel and perpendicular
pressures for each species defined with respect to the magnetic field direction ($\mathbf{b}$%
). In the following they are assumed to be suitably prescribed
scalar fields, for example, to be considered as specified by means
of an appropriate fluid or kinetic closure condition or directly
determined via experimental observations or from numerical
experiments.
Then $\nabla \cdot \underline{%
\underline{\Pi }}_{s}=\nabla P_{s}+\mathbf{bB\cdot \nabla }\left( \frac{%
P_{s\parallel }-P_{s}}{B}\right) +(P_{s\parallel
}-P_{s})\mathbf{b\cdot \nabla b,}$ where invoking again Ampere's
law and neglecting finite-$\beta $
effects, one obtains $\mathbf{b\cdot \nabla b\cong }\nabla \ln B\mathbf{%
\cdot }\left(
\underline{\underline{\mathbf{1}}}-\mathbf{bb}\right) .$ \ In the
following we shall ignore the effects due to the gravitational
self-field of the plasma and require
\begin{equation}
P_{s\parallel }=P_{s}+\delta _{1}p_{s},  \label{A-1}
\end{equation}%
\
\begin{equation}
\mathbf{V}_{s}(\mathbf{r},t)=V_{Ts}\mathbf{e}_{\varphi }+\delta _{2}V_{R}%
\mathbf{e}_{R}+\delta _{2}V_{z}\mathbf{e}_{z},  \label{A-2}
\end{equation}%
\begin{equation}
\mathbf{B}=\mathbf{B}_{p}+\varepsilon \mathbf{B}_{T}\equiv \nabla
\psi \times \nabla \varphi +\delta _{3}I(\psi )\mathbf{e}_{\varphi
},  \label{A-3}
\end{equation}%
where
\begin{equation}
\delta _{1},\delta _{2},\delta _{3}\ll 1,  \label{A-4}
\end{equation}%
where $\delta _{i}$ ($i=1,2,3$) are dimensionless parameters. Here
the notation
is standard. \ In particular $\mathbf{e}_{\varphi }=R\nabla \varphi ,$ $%
\mathbf{e}_{R}$ and $\mathbf{e}_{z}$ are the toroidal, radial and $z$%
-direction unit vectors, $\Omega _{s}$ is the toroidal angular
velocity of each species and $\psi $ is a suitable poloidal flux function. Moreover, $%
V_{Ts}\equiv \Omega _{s}R,$ $V_{R}$ and $V_{z}$ are the toroidal,
radial and
$z$-direction flow velocities, while $\mathbf{B}_{p}$ and $\delta _{3}%
\mathbf{B}_{T}$ denote the poloidal and toroidal magnetic fields.

\section{G-Hall-MHD equilibria with a purely poloidal magnetic field}

Let us now investigate in particular the case in which the radial and $z$%
-direction flow velocities and the toroidal magnetic field are
negligible while the fluid pressure of each species is isotropic.
In this case the gravitational-Hall-MHD (G-Hall-MHD) formulation
is simply achieved by invoking the stationary Maxwell equations
for the equilibrium EM fields and the momentum equations for
$\mathbf{V}_{s}$ for each species. Let us determine explicitly the
toroidal angular rotation velocity $\Omega _{s}$ for each species.
In this case, invoking the ordering (\ref{A-4}) and neglecting
corrections proportional to $\delta _{i}$ (for $i=1,2,3$), the
$\nabla R$ component of
the momentum equation for ions gives \emph{the bifurcated solutions}:%
\begin{equation}
\Omega _{i}{}^{\pm }=\frac{1}{2m_{i}}\left\{ -q_{i}\frac{\left(
\nabla \psi \cdot \nabla R\right) }{cR}\pm \sqrt{\left(
q_{i}\frac{\left( \nabla \psi
\cdot \nabla R\right) }{cR}\right) ^{2}+4\frac{m_{i}}{R}\left[ \frac{1}{ni}%
\nabla _{R}P_{i}+q_{i}\nabla _{R}\phi +m_{i}\nabla _{R}U^{eff}\right] }%
\right\} ,  \label{ion toroidal rotation frequence}
\end{equation}%
where $q_{i}$ is the ion electric charge (while in the following
$q_{e}$ denotes the electron charge). For electrons, neglecting
electron-inertia effects and considering again the $\nabla R$
component
of the momentum equation, one obtains%
\begin{equation}
\Omega _{e}\cong c\frac{\partial \phi }{\partial \psi }+c\frac{1}{q_{e}n_{e}}%
\frac{\partial P_{e}}{\partial \psi }. \label{electron frequency}
\end{equation}%
Finally, in the same approximation, the $\nabla z$ components of
the same equations gives respectively two constraint equations for
the (isotropic) scalar pressure of each species
$P_{s}=n_{s}T_{s}$, i.e., respectively for ions
and electrons%
\begin{equation}
\nabla _{z}P_{i}+n_{i}q_{i}\left( -\nabla _{z}\phi +\frac{R\Omega _{i}}{cR}%
\left( \nabla \psi \cdot \nabla z\right) \right) +n_{i}m_{i}\nabla
_{z}U^{eff}=0.  \label{Pz}
\end{equation}%
\begin{equation}
\nabla _{z}P_{e}+n_{e}q_{e}\left( -\nabla _{z}\phi +\frac{R\Omega _{e}}{cR}%
\left( \nabla \psi \cdot \nabla z\right) \right) =0.  \label{Pze}
\end{equation}%
These results show at once that in the present case:

\begin{itemize}
\item The fluid
(G-Hall-MHD) equilibria admit generally two distinct solutions
$\Omega _{i}{}^{\pm }$ for the ion angular velocity. Both of them
are in principle admissible. Nevertheless, one can see that in the
limit in which all the terms become small and negligible except
the gravitational one, one recovers the customary Keplerian (or
pseudo-Keplerian) solution. Nevertheless, in general AD plasmas,
both signs are admissible and they correspond to the two possible
rotational directions of the ion species (clockwise or
counterclockwise, according to the unit vector which defines the
reference system). As a basic implication, in the two cases the
parallel and the diamagnetic currents may differ significantly. In
such a case the bifurcation is expected to give rise to quite
disparate physical conditions for the AD, which correspond
respectively to the occurrence of "strong" and "weak" poloidal
magnetic fields in the AD plasma.
\end{itemize}

The existence domain of the velocity solution is determined by
requiring the
strictly positivity of the argument of the square root:%
\begin{equation}
\left( q_{i}\frac{\left( \nabla \psi \cdot \nabla R\right)
}{cR}\right) ^{2}+4\frac{m_{i}}{R}\left[ \frac{1}{n_{i}}\nabla
_{R}P_{i}+q_{i}\nabla _{R}\phi +m_{i}\nabla _{R}U^{eff}\right] >0
\end{equation}%
This domain is determined by the magnetic field, the pressure, the
density, the electrostatic and gravitational profiles. Whenever
the above condition is violated, the ion velocity becomes
imaginary. As this is a non-physical solution, we say that in the
domains where this happens the equilibrium does not exist.

\begin{itemize}
\item The treatment of the gravitational field produced by the central
massive object is handled by the introduction of a pseudo-Newtonian potential $%
G^{eff}$ \cite{Cremaschini2008}$.$ This allows us to take into
account relativistic corrections carried by the local space-time
curvature.

\item The density and pressure profiles for ions and electrons ($n_{i},n_{e}$
and $P_{i},P_{e}$) remain in principle arbitrary, being subject
only to the species constraint equations (\ref{Pz}). In
particular, if the density profiles are prescribed together with
the pressure of each species on the equatorial plane, [i.e.$,$
$P_{s}(R,z=0)]$, Eq.(\ref{Pz}) determines uniquely, for each
species $s,$ the vertical profile (i.e., the $z-$dependence) of
the its partial pressure $P_{s}$.

\item In the central region of the AD where $\Omega _{e}$ is sufficiently
large [see Eq.(\ref{electron frequency})] the electrostatic field,
in particular its component $E_{\psi }=-\nabla \psi \cdot \nabla
\phi /|\nabla \psi |$ which drives the toroidal rotation of the
electrons, is expected to become very strong.
\end{itemize}

Let us now investigate the relevant Maxwell equations: the Poisson
equation for the electrostatic field $\mathbf{E}$ and Ampere's law
for the magnetic field. These give respectively for the
electrostatic potential $\phi $ and the poloidal flux function the
pde's
\begin{equation}
\frac{1}{R}\frac{\partial }{\partial R}\left( R\frac{\partial \phi }{%
\partial R}\right) +\frac{\partial ^{2}\phi }{\partial z^{2}}=-4\pi \left(
|q_{i}|n_{i}-|e|n_{e}\right) ,
\end{equation}%
\begin{equation}
\frac{1}{R}\frac{\partial ^{2}\psi }{\partial z^{2}}+\frac{1}{R}\frac{%
\partial ^{2}\psi }{\partial R^{2}}=-\frac{4\pi R}{c}\left(
|q_{i}|n_{i}\Omega _{i}-|e|n_{e}\Omega _{e}\right) ,
\end{equation}%
to be identified respectively with the Poisson equation and the
generalized Grad-Shafranov equation.

\section{Conditions of validity and closure conditions}

We now make a few comments on the theory developed here. The first
one concerns the assumption of a weakly collisional plasma for the
AD, whereby the various (plasma) species are allowed to have
different velocities. This requires that the mean free path for
all of the species must be assumed to be much larger than the
characteristic macroscopic scale length ($L$) of the relevant
fluid fields ($\phi _{1},...,\phi _{N}$). In particular, $L$ can
be identified locally with $L=\inf_{i=1,N}(L_{i}),$ where
$L_{i}^{-1}=\left\vert \nabla \ln \phi _{i}\right\vert $ is the
scale length associated with the gradient of the fluid field $\phi
_{i}.$ This assumption has the important consequence that the
current density (and hence the self-consistent magnetic field
which it generates) can become very strong even in a low density
plasma. A second aspect concerns the physical conditions in which
Hall-current effects are expected to play a significant role for
G-Hall-MHD
equilibria. In this regard it is important to point out again that Eq.(\ref%
{electron frequency}) implies that in the central regions of the
AD, where $\Omega _{e}$ is sufficiently large, the electrostatic
field can be expected to become very strong. Hence, local
deviations from quasi-neutrality, should be expected in these
equilibria. This means that the Debye length $\lambda
_{Ds}=1/\sqrt{\frac{4\pi Z_{s}^{2}e^{2}n_{s}}{T_{s}}}$ of each
species may become locally comparable to the macroscopic
characteristic scale length ($L$) of the plasma equilibrium. The
third aspect concerns the peculiar feature discovered here for
G-Hall-MHD equilibria, related to the existence of bifurcated
solutions for the ion toroidal angular velocity. It is obvious
that, for a prescribed magnetic field (produced by the central
object), the direction of ion flow velocity in the outer regions
of the AD uniquely determines the sign of the ion toroidal angular
frequency also in the central region. As a consequence, it depends
on the direction of injection for the charged particles at the
external boundary of the disk, where the plasma flow is expected
to be almost Keplerian. With the specification of this boundary
condition the equilibrium solution therefore becomes unique.

Another issue is related to the closure conditions for the
Hall-MHD equations. \ To resolve the indeterminacy in the density
and pressure of each species the equilibrium magnetic and electric
fields have been determined under the assumption of suitably
prescribing the pressure of each species on the equatorial plane
$P_{s}(R,z=0).$ We can motivate the necessity and the use itself
of the closure relations by the following arguments:

1) this is a standard approach in plasma physics. For example, the
same technique is used while studying plasma equilibria for
laboratory plasmas;

2) generally, the set of fluid equations for plasmas may not be
closed even at equilibrium, at least in the case of locally
non-Maxwellian kinetic equilibria.

In general, the closure problem may be solved by adopting one of
the following alternative approaches:

1) taking the density or temperature profiles for each species
from experimental data. This would require observational data to
be reduced and compared with theoretical models.

On the other hand, the theoretical model also suggests possible
interesting developments related specifically to the adoption of:

2) phenomenological fluid closure conditions (from numerical
experiments);

3) kinetic closure conditions, in which the undetermined fluid
fields are prescribed via kinetic theory.

Finally - as indicated above - a completely different viewpoint
lies in the investigation of kinetic equilibria. This involves the
requirement that the G-Hall-MHD equilibrium corresponds actually
to an underlying kinetic equilibrium, to be suitably prescribed.
This point and related issues relevant for the fluid/kinetic
evaluation of G-Hall-MHD equilibria will be
discussed in greater detail in the accompanying paper \cite{Cremaschini2008b}%
.

\section{Conclusions}

In this paper the relevant equations which describe G-Hall-MHD
equilibria have been investigated. From the above analysis we
summarize the following main aspects:

1) the relevant equations describing fluid equilibria of this type
have been obtained under the assumption of axial symmetry and a
collisionless plasma;

2) the effect produced by equilibrium diamagnetic currents
generated by the plasma, including non-neutrality and the Hall
current, have been taken into account;

3) the effect of differential rotation produced by the non rigid
toroidal rotation of the AD has been taken into account by
including the toroidal angular frequency of each species;

4) allowance has been made for the inclusion of relativistic
effects by taking into account the space-time curvature generated
by the central compact object;

5) no assumption of local quasi-neutrality has been invoked.

The present theory can be generalized in several ways, to include
- in particular - the kinetic treatment of G-Hall-MHD equilibria
(to be discussed in the accompanying paper
\cite{Cremaschini2008b}), species-dependent flow velocities having
both toroidal and radial flow velocities, as well as the
additional presence of a toroidal magnetic field produced by the
plasma flows, of an anisotropic pressure tensor and of
self-gravitating AD's. Based on the present theory, G-Hall-MHD
equilibria can be investigated utilizing a perturbative solution
method based on a power series expansion near to the equatorial
plane ($z=0$) \cite{Cremaschini2008}. This permits an analysis of
the combined role of Hall, diamagnetic and relativistic curvature
effects, which are all expected to play a significant role in the
central regions of AD's. \

\bigskip

\section*{Acknowledgments}
This work has been developed in cooperation with the CMFD Team,
Consortium for Magneto-fluid-dynamics (Trieste University,
Trieste, Italy), within the framework of the MIUR (Italian
Ministry of University and Research) PRIN Programme: {\it Modelli
della teoria cinetica matematica nello studio dei sistemi
complessi nelle scienze applicate}. Support is acknowledged (by
A.B.) from ICTP (International Center for Theoretical Physics,
Trieste, Italy) and the University of Trieste, Italy, (by M.T.)
from COST Action P17 (EPM, {\it Electromagnetic Processing of
Materials}) and GNFM (National Group of Mathematical Physics) of
INDAM (Italian National Institute for Advanced Mathematics).

\section*{Notice}
$^{\S }$ contributed paper at RGD26 (Kyoto, Japan, July 2008).

\end{document}